\documentclass[12pt,letterpaper]{article}
\usepackage[latin1] {inputenc} 
\usepackage{amsmath} \usepackage{amsfonts} \usepackage{amssymb} 
\begin{document}

\title{First Order Actions and Duality}
\author{Alejandro Gaona\footnote{email:gaona@nucleares.unam.mx}\ \  and J. Antonio Garc\'\i a\footnote{email:
garcia@nucleares.unam.mx} \\
\em Instituto de Ciencias Nucleares, \\
\em Universidad Nacional Aut\'onoma de M\'exico\\
\em Apartado Postal 70-543, M\'exico D.F., M\'exico}
\maketitle
\abstract{ We consider some aspects of classical S-duality transformations in first order actions taken into 
account the general
covariance of the Dirac algorithm and the transformation properties of the Dirac bracket. By classical S-Duality 
transformations 
we mean a field redefinition that interchanges the
equations of motion and its associated Bianchi identities. By working from a first order variational principle and
performing the corresponding Dirac analysis we find that the standard electro-magnetic duality can be 
reformulated as a
canonical local transformation. The reduction from this phase space to the original phase space variables coincides with the
well known result about duality as a canonical non local transformation. We have also applied our ideas to the bosonic string. These Dualities are
not canonical transformations for the Dirac bracket and relate actions with different kinetic 
terms in the reduced space. }

\section{Introduction}

The word ``Duality'' is ubiquitous in recent literature about string theory but as is well known duality transformations
have its origin in Statistical Mechanics and Conformal Field Theories as a consequence of locality and associativity of
the operator product expansion. In this sense they are symmetries of some partition functions or symmetries of the
spectra in some quantum field theories, but the analysis of to what extent or in what sense are these transformations
also symmetries of the underlaying variational principle (when we have one at hand) is less understood.   
Within the context of Statistical Mechanics \cite{Savit} the idea of duality transformation relates a system described
by 
a Hamiltonian in different coupling regimen. An important application of this idea was the determination of the exact
temperature at which the phase transition of two-dimensional Ising model takes place. The key point here is that the two
models are related by a field redefinition between the spin variables, the link variables associated with the
corresponding lattice and the new dual spin variables in the so called Dual lattice.

Another interesting application of the idea of duality transformation in field theory is the celebrated duality between
the Sine-Gordon and Thirring model in 1+1 dimensions. Here the interesting point is that perturbative solutions of one
model are nonperturbative solutions of the other model. From the point of view of symmetry mappings
(transformations that leave the solution space of a given variational principle invariant) these type of transformations
are also symmetries in the sense that map the space of solutions of a given problem into itself. Indeed, these
transformations connect solutions from the weak coupling regime to the strong coupling one. These type of
transformations are called S-duality which consist in a (conjectural) symmetry relating the strong coupling regime of
one theory with the weak coupling limit of the same or other theory. These S-dualities are generalizations of the
original conjecture by Montonen and Olive \cite{Montonen-Olive}.  

The non-abelian S-duality case is also very interesting. As an example, consider the principal chiral model with group
$SU(2)$. In this case it is also possible to construct a canonical  map of this model to a theory which
turns out to be the non abelian dual with respect to the left action of the whole group \cite{Gaume, CZ, CZ1}. Here also the
canonical mapping is performed at the level of the reduced space (namely we solve the constraint and then map the
original system to the new one). 

In string theory and 2-dimensional conformal field theory, duality is an important tool to show the equivalence of
different geometries and/or topologies \cite{nn}. 
In this context duality transformations were first described in the context of toroidal compactifications, also known as T-Duality maps
(see for example \cite{p} and references therein). For the simplest case of a single compactified dimension of radius $R$, the spectra of the interacting theory
is left unchanged under the replacement $R\to\alpha'/R$ provided one also transform the dilaton field $\phi\to\phi-\log
(R/\sqrt{\alpha'})$ in the effective string action. This simple case can be generalized to arbitrary toroidal
compactifications described by a constant metric $g_{ij}$ and antisymmetric tensor $B_{ij}$.

The aim of this letter is twofold. From one hand to consider the S-duality transformation 
as a mapping between two classical theories. We will analyze the case were this map is a symmetry of the equations of motion but in general is not 
a symmetry of the associated variational principle. 
Starting from S-duality as a field redefinition (that can be non
local), {\em i.e.}, a transformation that map a given theory to a new one with the same physical content but using different
``names'' for its description 
at a classical level (but that could produce different physics at quantum level) we
will answer the question of to what extent or in what sense are this duality map a symmetry of a given
variational principle. 

From the other hand we want to analyze the problem of how a canonical non-local transformation that is a symmetry of the equations of motion can be reformulated in a bigger space (with some auxiliary fields added)  as a canonical local transformation. 

In section 2 we will review the basic arguments to implement this duality as non local canonical transformation 
and discuss why these symmetries can not be implemented as symmetries of the variational principle without solving the 
constraints. In Section 3 we will review the effects that such transformations produce in the Dirac constraint
analysis and in particular in the Dirac Bracket. In Section 4 we will
present an interesting example by formulating the bosonic string action as a first order variational principle and show
the general transformation properties of the system under the analog of these S-Duality transformations.  
Section 5 is devoted to conclusions and perspectives.   

\section{Duality as a symmetry of the first order variational principle}

As a warm up consider the first order action in $2n$ dimensions defined by
\begin{equation}
\label{action-d} S=-\frac{1}{2n}\int d^{2n}x\, F_{\mu_1...\mu_n}F^{\mu_1...\mu_n}-
F^{\mu_1...\mu_n}\partial_{\mu_1}A_{\mu_2...\mu_n}.
\end{equation}
Here $F$ and $A$ are independent variables. That this action and the original source free Maxwell action in $2n$
dimensions are equivalent is a consequence of the fact that $F$ are auxiliary fields in the action (\ref{action-d}).
The definition of the field strength in terms of the $(n-1)-$form  $A$ is $F_{\mu_1\mu_2....\mu_n}=\partial_{[\mu_1}
A_{\mu_2....\mu_n]}$. 

The equations of motion that follow from the action (\ref{action-d}) are
\begin{equation}
\partial_{\mu_1} F^{\mu_1...\mu_n}=0,\qquad F_{\mu_1...\mu_n}-
\partial_{[\mu_1} A_{\mu_2...\mu_n]}=0.
\end{equation}
We want to consider the behavior of the action and the equations of motion under the transformation  
\begin{equation}
\label{field-redef} F\to \ ^*F, \qquad A\to A.
\end{equation}
The dual of the field $F$ is defined by
\begin{equation}
^* F^{\mu_1....\mu_n}=\frac{1}{n!}\varepsilon^{\mu_1...\mu_{n}\mu_{n+1}....\mu_{2n}} F_{\mu_{n+1}....\mu_{2n}},
\end{equation}
where $\varepsilon$ is the Levi-Civita symbol with $\varepsilon^{01...2n}=1$ in $2n$ dimensions. Taking into account that
double duality has the property \cite{deser}
\begin{equation}
^{**} F=F, \quad n=2k+1, \qquad ^{**} F=-F, \quad n=2k,
\end{equation}
and that
\begin{equation}
\label{alld}
F^2=-^*F^2,
\end{equation}
is valid in all dimensions, the new transformed (or redefined) action is
\begin{equation}
\label{action-dual} S_D=-\frac{1}{2n}\int d^{2n}x\,\left( -F_{\mu_1\mu_2...\mu_n}F^{\mu_1\mu_2...\mu_n}-
\frac{1}{n!}\varepsilon^{\mu_1\mu_2...\mu_{2n}} F_{\mu_{n+1}....\mu_{2n}} 
\partial_{\mu_1}A_{\mu_2...\mu_n}\right),
\end{equation}
where we have used the transformation (\ref{field-redef}) and the relation (\ref{alld}).
The equations of motion that follow from this Dual action are
\begin{equation}\label{dual-em}
\varepsilon^{\mu_1\mu_2...\mu_{2n}}\partial_{\mu_1} F_{\mu_{n+1}....\mu_{2n}} =0,\qquad F_{\mu_{n+1}...\mu_{2n}}+\frac{1}{n!}\varepsilon_{\mu_1\mu_2...\mu_{2n}}  
\partial^{\mu_1}A^{\mu_2...\mu_n}=0.
\end{equation}
Is the transformation (\ref{field-redef}) a symmetry of the first order action (\ref{action-d})? We observe that if we use the second equation of motion in (\ref{dual-em}) and substitute it on the first equation in (\ref{dual-em}) the resulting  equations of motion are the Maxwell equations in $2n$ dimensions. The same is true for the equations of motion of the original action (\ref{action-d}).
We conclude that in the configuration space (the space defined by $A$) the two actions have the same equations of motion. Nevertheless the first order actions and its first order equations of
motion are quite different. They define different symplectic structures and as a consequence they differ by a
non-canonical transformation (the field redefinition (\ref{field-redef})). In phase space, upon the elimination of the
auxiliary variables $F$ by means of its own equations of motion, we recover the well known S-duality implemented as a  non-local canonical
transformation \cite{lozano}. 

After reviewing some basic facts about the behavior  of the Dirac algorithm and the Dirac Bracket under field redefinitions we will show that in the phase space defined by the variables $A,F,\pi_A,\pi_F$ where $\pi_A$ and $\pi_F$ are the conjugate momenta to $A$ and $F$ respectively, the transformation (\ref{field-redef}) can be implemented as a local canonical transformation.

\section{Dirac formalism and Field Redefinitions}

\subsection{Dirac algorithm}

We will review here some ideas about the behavior of the Dirac algorithm under canonical
transformations.
 
Consider and extended first order Lagrangian in $d$-dimensions whose extended action is (for details see \cite{Han})
\begin{equation}
S_E=\int d^d x \left(\dot\phi \pi -H_c(\phi,\pi)-\lambda^a\gamma_a(\phi,\pi)-
\lambda^\alpha\chi_\alpha(\phi,\pi)\right),
\end{equation}
where $\phi(x)$ are some fields that define the theory and $\pi(x)$ its associated momenta. $H_c$ is the canonical
Hamiltonian and $\gamma_a\approx 0$ and $\chi_\alpha\approx 0$ are the first class constraints and the second class
constraints respectively. $\lambda^a$ and $\lambda^\alpha$ are Lagrange multipliers. The extended action has been
obtained after performing a complete Dirac analysis of the theory. We suppose that the theory is consistent and regular
and that there are no more constraints that the ones that we have written in the extended variational principle.

The Poisson structure is
\begin{equation}
\label{c-bra}
\{\phi(x),\pi(y)\}=\delta(x-y),
\end{equation}
and the associated Dirac bracket 
\begin{equation}
\label{d-bra}
\{A(x),B(y)\}^*=\{A(x),B(y)\}-\int d\eta d\xi\,
\{A(x),\chi_\alpha(\eta)\}C^{\alpha\beta}(\eta,\xi)\{\chi_\beta(\xi),B(y)\},
\end{equation}
where
\begin{equation}
C_{\alpha\beta}(x,y)=\{\chi_\alpha(x),\chi_\beta(y)\},
\end{equation}
and $C^{\alpha\beta}$ denotes the inverse of the matrix $C_{\alpha\beta}$.

The algebra of constraints is\footnote{Up to terms quadratic in second class constraints.}
\begin{eqnarray}
\{\gamma_a(x),\gamma_b(y)\}=\int dz\,C_{ab}^c(x,y,z)\gamma_c(z),\\
\{\gamma_a(x),\chi_\alpha(y)\}=\int dz\,\left( C_{a\alpha}^b(x,y,z)\gamma_b(z)+ C^\beta_{a\alpha}(x,y,z)\chi_\beta(z)\right),\\
\{H_c, \gamma_a(x)\}=\int dy\, V^b_a(x,y)\gamma_b(y),\\
\{H_c,\chi_\alpha(x)\}=\int dy\,\left( V_\alpha^a(x,y)\gamma_a(y)+ V^\beta_\alpha(x,y)\chi_\beta(y)\right).
\end{eqnarray}
Now suppose that we implement the canonical transformation given by
\begin{equation}\label{ct}
\phi\to \Phi(\phi,\pi),\qquad \pi\to \Pi(\phi,\pi).
\end{equation}
In general, such transformation will change the structure functions of the algebra of constraints and/or the structure of
the Dirac algorithm. It could change also the functional structure of the constraints. Under general assumptions (regularity
conditions) the constraint surface will change to another surface with the same geometrical content, {\em i.e}, it is
described by the same number of first and second class constraints. So the rank of the matrix formed by the
Poisson brackets of all the constraints between themselves, remain constant under the canonical transformation. We will
suppose here that this is the case. It is worth noticing that the canonical transformation is defined for
the standard symplectic structure (\ref{c-bra}). Then we expect that, in general, the Dirac bracket (\ref{d-bra}) will
be modified by the transformation. An observation that will be of relevance in what follows is that such canonical transformation in phase
space in not canonical with respect to the Dirac bracket (\ref{d-bra}). This is precisely the trick used to
construct a ``canonical representation of the constraint surface'' \cite{PG} where the first class constraints are
realized as a subset of momenta and the second class constraints as a proper subset of fields and its associated momenta. In
these coordinates the Dirac bracket takes the standard form of an ordinary Poisson structure in Darboux coordinates.

Under the canonical transformation (\ref{ct}) the new Dirac bracket will take the form
\begin{eqnarray}\nonumber
\{A(x),B(y)\}_{(\Phi,\Pi)}^*=\{A(x),B(y)\}_{(\Phi,\Pi)}\\
 -\int d\eta d\xi \{A(x),X_\alpha(\eta)\}_{(\Phi,\Pi)}{\cal C}^{\alpha\beta}(\eta,\xi)\{X_\beta(\xi),B(y)\}_{(\Phi,\Pi)},
\end{eqnarray}
where all the Poisson brackets are calculated with respect to the new variables $\Phi,\Pi$. The new constraints
$X^\alpha(\Phi(\phi,\pi),\Pi(\phi,\pi))=\chi^\alpha(\phi,\pi)$ and ${\cal C}_{\alpha\beta}$ are defined by
\begin{equation}
\label{matrix-c} {\cal C}_{\alpha\beta}(x,y)=\{X^\alpha(x),X^\beta(y)\}_{(\Phi,\Pi)}.
\end{equation}
In particular a regular field redefinition $\phi\to \Phi(\phi)$ can always be extended to a canonical transformation \cite{gold}. The canonical transformation associated with this field redefinition
is  generated by a type 2 generating function given by $F_2=F(\phi)\Pi$
\begin{equation}
\pi(x)=\frac{\delta F_2}{\delta \phi(x)},\qquad \Phi(x)=\frac{\delta F_2}{\delta \Pi(x)}.
\end{equation}
After the implementation of the second class constraints, by solving them, we
arrive to a partially reduced variational principle with the general structure
\begin{equation}
\label{action-red} S_R=\int d^d x\, \dot\xi^r \ell_r(\xi) -H_R(\xi)-\lambda^a\gamma_a(\xi),
\end{equation}
where $\xi^r(x)$ denotes the remaining fields and momenta that are the coordinates of the reduced space. The reduced
Dirac bracket is the inverse of the symplectic structure defined by the first order action (\ref{action-red})
\begin{equation}
\{A(x),B(y)\}^*_R=\sigma^*_R(x,y),
\end{equation}
where
\begin{equation}
\eta_{rs}(x,y)=\frac{\delta\ell_s(x)}{\delta \xi^r(y)}-\frac{\delta\ell_r(y)}{\delta \xi^s(x)},
\end{equation}
and $\eta^{-1}(x,y)=\sigma^*_R(x,y)$.

\subsection{Dirac Bracket and Field redefinitions}

Consider a general Poisson structure defined by the bracket
\begin{equation}
\label{orig-bra}
\{A(z),B(z)\}=\frac{
\partial A}{
\partial z^a}\sigma^{ab}(z)\frac{
\partial B}{
\partial z^b},
\end{equation}
in a $n$-dimensional space defined by the variables 
$z^a, a=1,2...n$. Now, lets perform a ``field redefinition'' given by
\begin{equation}
\label{z-Z} z^a=Z^a(z),
\end{equation}
where the functions $Z^a$ are independent and regular so the inverse transformation exist and is also regular. The
redefined bracket is
\begin{equation}
\{{\cal A}(Z),{\cal B}(Z)\}=
\frac{
\partial {\cal A}}{
\partial Z^c}\frac{
\partial Z^c}{
\partial z^a} \sigma^{ab}(z)\frac{
\partial Z^d}{
\partial z^b} \frac{
\partial {\cal B}}{
\partial Z^d},
\end{equation}
where ${\cal A}(Z)=A(z(Z))$ and the same for ${\cal B}$. Define
\begin{equation}
\label{sigma-transf}
\Sigma^{cd}(Z)\equiv\frac{
\partial Z^c}{
\partial z^a} \sigma^{ab}(z)\frac{
\partial Z^d}{
\partial z^b},
\end{equation}
so the tensor $\sigma^{ab}$ transform as a second rank tensor. This definition is natural if we observe that 
$\{Z^a,Z^b\}_Z=\Sigma^{ab}(Z)$. In practice we calculate the function in the right of (\ref{sigma-transf}) and read out the result with a
 simple rename  of the variable $z$ for $Z$. Then the new bracket in the variables $Z$ is
\begin{equation}
\{{\cal A}(Z),{\cal B}(Z)\}=\frac{
\partial \cal A}{
\partial Z^a}\Sigma^{ab}(Z)\frac{
\partial \cal B}{
\partial Z^b}.
\end{equation}
Now we will apply this transformation rule to the case of the Dirac Bracket. The
Dirac Bracket can be written as
\begin{equation}
\{A(z),B(z)\}^*=\frac{
\partial A}{
\partial z^a}{\sigma^*}^{ab}(z)\frac{
\partial B}{
\partial z^b},
\end{equation}
where
\begin{equation}
{\sigma^*}^{ab}(z)={\sigma}^{ab}(z)-{\sigma}^{ac}(z)
\frac{
\partial \chi_\alpha(z)}{
\partial z^c} C^{\alpha\beta}(z)
\frac{
\partial \chi_\beta(z)}{
\partial z^d}{\sigma}^{db}(z).
\end{equation}
Here $\chi_\alpha(z)=0$ are the set of second class constraints and the matrix $C_{\alpha\beta}$ is defined by
(\ref{matrix-c}).

Then the transformed Dirac bracket is
\begin{equation}
\{{\cal A}(Z),{\cal B}(Z)\}=\frac{
\partial \cal A}{
\partial Z^a}{\Sigma^*}^{ab}(Z)\frac{
\partial \cal B}{
\partial Z^b},
\end{equation}
where
\begin{equation}
{\Sigma^*}^{cd}(Z)=\frac{
\partial Z^c}{
\partial z^a} {\sigma^*}^{ab}(z)\frac{
\partial Z^d}{
\partial z^b},
\end{equation}
or in terms of the Poisson structure of the base space $\Sigma^{ab}$ 
\begin{equation}
{\Sigma^*}^{ab}(Z)={\Sigma}^{ab}(Z)-{\Sigma}^{ac}(Z)
\frac{
\partial X_\alpha(Z)}{
\partial Z^c} {\cal C}^{\alpha\beta}(Z)
\frac{
\partial X_\beta(Z)}{
\partial Z^d}{\Sigma}^{db}(Z),
\end{equation}
where $X_\alpha(Z)=\chi_\alpha(z(Z))$ and 
\begin{equation}
{\cal C}_{\alpha\beta}(Z)=\frac{
\partial X_\alpha(Z)}{
\partial Z^a}{\Sigma}^{ab}(Z)\frac{
\partial X_\beta(Z)}{
\partial Z^b}.
\end{equation}
 ${\cal C}^{\alpha\beta}$ is the inverse of ${\cal C}_{\alpha\beta}$. In particular if the field redefinition
(\ref{z-Z}) is a canonical transformation  for the original bracket then
\begin{equation}
\Sigma^{ab}=\sigma^{ab}.
\end{equation}
In this case $C_{\alpha\beta}(z)={\cal C}_{\alpha\beta}(Z)$. It is clear that
this canonical transformation is not a canonical transformation for the Dirac bracket (\ref{orig-bra}). Indeed, from $\sigma^{ab}=\Sigma^{ab}$ we can not deduce that $\sigma^*=\Sigma^*$.

\section{First order Maxwell action}

In this section we will apply the ideas of the previous sections to the first order action (\ref{action-d}) in $d=4$. Lets start from the first order action
\begin{equation}
\label{EM-action} S_0(F,A)=\frac{1}{2g^2}\int d^4x\, (\frac12 F^2-F^{\mu\nu}\partial_{[\mu}A_{\nu]}),
\end{equation}
where $g$ is the coupling constant. This action can be rewritten as
\begin{equation}
S_0(F,A)=\frac{1}{2g^2} \int d^4x\, ( - F^{0i}{\dot A}_i+\frac12 F^2-F^{ij}\partial_{[i}A_{j]}-F^{0i}\partial_iA_0).
\end{equation}
The Lagrangian action in phase space is 
\begin{equation}
S_0(F^{\mu\nu},A^\mu,\pi^{\mu\nu},\pi_\mu)=\int d^4x\,(\pi_\mu{\dot A}^\mu+{\dot F}^{\mu\nu}\pi_{\mu\nu}-H_c),
\end{equation}
where
\begin{equation}
H_c=\frac{1}{2g^2}(-\frac12 F^2+F^{ij}
\partial_{[i}A_{j]}+F^{0i}
\partial_iA_0),
\end{equation}
is the canonical Hamiltonian.

The first class constraints are
\begin{equation}
\gamma_0=\pi_0,\qquad \gamma=
\partial^i\pi_i,
\end{equation}
and the second class constraints 
\begin{equation}
\chi_{1(i)}=\pi_{0i},\quad \chi_{2(ij)}=\pi_{ij},\quad \chi_{3(j)}=\pi_j+\frac{1}{2g^2}F^{0j},
\end{equation}
where
\begin{equation}
{\chi_4}_{(kl)}=F_{kl}-
\partial_{[k}A_{l]}.
\end{equation}
The number of first class constraints is 2 and for the second class is 12, so the number of degrees of freedom is 2 as
expected.

The Dirac brackets that are different from zero are
\begin{eqnarray}
\{F^{ij}(x),F_{0k}(y)\}^{*}&=&-2g^{2}\delta^{[i}_{k}\partial^{j]}\delta(x-y),\\
\{A^{i}(x),F_{0k}(y)\}^{*}&=&2g^{2}\delta^{i}_{k}\delta(x-y),\\
\{A^{\mu}(x),\pi_{\nu}(y)\}^{*}&=&\delta^{\nu}_{\mu}\delta(x-y),\\
\{F^{ij}(x),\pi_{k}(y)\}^{*}&=&\delta^{[i}_{k}\partial^{j]}\delta(x-y).
\end{eqnarray}

The reduction, by solving the second class constraints, is straightforward and the result is the standard Maxwell action
in phase space. The reduced Dirac bracket coincides with the standard bracket associated to the Darboux symplectic
structure of the phase space defined by $A^\mu$ and its momenta $\pi_\mu$. Another reduction is also possible to recover
the first order variational principle (\ref{EM-action}). This reduction is performed by the elimination of the canonical
momenta using the equations of motion for the Lagrangian multipliers and the momenta. The relevant reduced Dirac bracket
is
\begin{equation}
\{A^i(x),F_{0j}(y)\}^*_R=2g^2\delta^i_j\delta(x-y),
\end{equation}
as expected from the analysis of the previous section.

Now lets consider the Dual theory. This theory can be obtained from the previous one by performing the field
redefinition
\begin{equation}
\label{dual-transform} F\to {\cal F}=\frac{1}{g^2}\, {^*F},\qquad A\to g^2\Lambda,
\end{equation}
The dual theory is
\begin{equation}
\label{EM-dual} S_D({\cal F}^{\mu\nu},\Lambda^\mu)=\frac{g^2}{4}\int d^4x\,  (-{\cal
F}^2+\varepsilon^{\alpha\beta\mu\nu}{\cal F}_{\mu\nu}
\partial_{[\alpha}\Lambda_{\beta]}).
\end{equation}
By extending the point transformation (\ref{dual-transform}) to all phase space as a canonical transformation with
respect to the standard Poisson bracket we find that the required transformation is generated by 
\begin{equation}
\label{can-gen} F_2(F_{\mu\nu}, A_\mu, \Pi^\mu,\Pi^{\mu\nu})=\frac{1}{g^2}\int d^3
x(\frac12\Pi^{\mu\nu}\varepsilon_{\mu\nu\sigma\rho}F^{\sigma\rho}+\Pi^\mu A_\mu).
\end{equation}
Explicitly the canonical transformation is given by
\begin{eqnarray}
\pi_\mu=\frac{\delta F_2}{\delta A^\mu}, \quad \pi_{\mu\nu}=\frac{\delta F_2}{\delta F^{\mu\nu}},\\
\Lambda^\mu=\frac{\delta F_2}{\delta \Pi_\mu};\quad 
{\cal F}^{\mu\nu}=\frac{\delta F_2}{\delta \Pi_{\mu\nu}},
\end{eqnarray}
or
\begin{eqnarray}
\label{canonic} 
\nonumber {\cal F}^{\mu\nu}=\frac{1}{2g^2}\varepsilon^{\mu\nu\sigma\rho}F_{\sigma\rho}, \qquad
\pi_{\mu\nu}=\frac{1}{2g^2}\varepsilon_{\mu\nu\sigma\rho}\Pi^{\sigma\rho},\\
 \Lambda^\mu= \frac{1}{g^2}A^\mu, \qquad \pi^\mu=\frac{1}{g^2} \Pi^\mu,
\end{eqnarray}
where $\varepsilon^{\mu\nu\sigma\rho}\varepsilon_{\mu\nu\sigma\rho}=-4!$ and $\Pi^\mu,\Pi^{\mu\nu}$ are the momenta
associated to the configuration variables $\Lambda_\mu,{\cal F}_{\mu\nu}$.  In the following we will consider the new
theory in terms of the variables $\Lambda_\mu,{\cal F}_{\mu\nu}$. This is the theory that we call the dual theory. The
mapping of the constraint surface is
\begin{equation}
\Gamma_0=\Pi^0,\qquad \Gamma=
\partial_i\Pi^i,
\end{equation}
and the new second class constraints are
\begin{equation}
X_{1(ij)}=\Pi_{ij},\quad X_{2(i)}=\Pi_{0i},\quad X^{3(i)}=\Pi_i-\frac{g^2}{4}\varepsilon_{ikl}{\cal F}^{kl},
\end{equation}
and
\begin{equation}
X^{4(k)}={\cal F}^{0k}-\frac12 \varepsilon^{ijk}
\partial_{[i}\Lambda_{j]}.
\end{equation}
Of course all these constraints can be obtained from the Dual action (\ref{EM-dual}) by performing  systematically the
Dirac analysis. As a consequence of the deformation of the second class constraint surface the Dirac bracket changes.

The dual Dirac brackets different from zero are
\begin{eqnarray}
\{\Lambda_{i}(x),\Pi^{j}(y)\}^{*}_{D}&=&\delta^{j}_{i}\delta(x-y),\\
\{\Lambda^{i}(x),\mathcal{F}^{jk}(y)\}^{*}_{D}&=&
\frac{2}{g^{2}}\epsilon^{ijk}\delta(x-y),\\
\{\mathcal{F}_{0k}(x),\mathcal{F}^{ij}(y)\}^{*}_{D}&=&
-\frac{2}{g^{2}}\delta^{i}_{[k}\delta^{j}_{l]}\partial^{l}\delta(x-y),\\
\{\mathcal{F}_{0k}(x),\Pi_{l}(y)\}^{*}_{D}&=&-\epsilon_{lkj}
\partial^{j}\delta(x-y).
\end{eqnarray}

 Notice that the Dual Dirac bracket is different from the Dirac bracket of the original theory. This will imply that the reduced theories
 will be related by a non-canonical transformation. To see it in detail lets reduce the Dual theory
by enforcing the second class constraints and solving them in such a way that the remaining variables will be $F$ and $A$. The
resulting reduced action is
\begin{equation}
\label{Fo-dual} S_D=\int d^4x\, ( \varepsilon_{ijk} F^{ij}{\dot A}^ki-\frac12 F^2-\varepsilon_{ijk}
F^{0k}\partial_{[i}A_{j]}-\varepsilon_{ijk}F^{ij}
\partial^kA_0),
\end{equation}
that correspond to the Dual first order action calculated directly from (\ref{EM-dual}) using the field redefinition
(\ref{canonic}) that correspond to a non-canonical transformation. So, the canonical transformation in phase space
projects, after the reduction, into a non-canonical transformation in the reduced space. The relevant reduced bracket is now
\begin{equation}
\{A^i(x),F_{jk}(y)\}^*_R=\varepsilon_{ijk}\delta(x-y),
\end{equation}
that coincides with the one obtained from the reduced Dual first order action using (\ref{Fo-dual}). 

The original action $S_0$ and the Dual action $S_D$ gives the same equations of motion in the configuration space
defined by $A$ but their equations of motion are different in the space defined by $F$ and $A$ as we have advanced at the end of section 2.

As in the original action we can also reduce the Dual action by solving the second class constraints for the variables
$A$ and its conjugate momenta. In that case we recover the analysis of \cite{lozano}, and the Duality transformation is
canonical and non-local. To see this consider the generator $F_2$ defined in (\ref{can-gen}) and project  it over the
constraint surface. The canonical generator is then
\begin{equation}
\label{proj-gen} F_1(\Lambda,A)=\int d^3x \varepsilon_{ijk}{\cal F}^{jk}(\Lambda) A^i=\int d^3 x{\cal F}_{0i}(\Lambda)
A^i=\int d^3 x \varepsilon^{ijk}
\partial_{[j}\Lambda_{k]}A_i,
\end{equation}
which coincides with  the generator used to construct the canonical non-local Dual transformation in \cite{lozano}. This
transformation is
\begin{equation}
\pi^{i}=\frac{\delta F_1}{\delta A^i}=\varepsilon^{ijk}
\partial_j\Lambda_k,\quad \Pi^{i}=-\frac{\delta F_1}{\delta \Lambda^i}=\varepsilon^{ijk}
\partial_jA_k.
\end{equation}
 The remaining first class constraints are
\begin{equation}
\gamma_0=\pi^0,\qquad \gamma=
\partial^i\pi_i= \frac{1}{2g^2}
\partial_i F^{0i}(A)=
\partial_i\varepsilon_{ijk} F^{jk}(A)  \equiv 0,
\end{equation}
for the original theory and
\begin{equation}
\Gamma_0=\Pi^0,\qquad \Gamma=
\partial^i\varepsilon_{ijk}{\cal F}^{jk}(\Lambda) =
\partial_i{\cal F}^{0i}(\Lambda)= 
\partial_i \varepsilon^{ijk}
\partial_{[j}\Lambda_{k]} \equiv 0,
\end{equation}
for the dual theory. So the Gauss law transform from the electric to the magnetic case, as expected and the reduction imply that the duality can only be implemented as a symmetry on-shell.

\section{First order Bosonic String action}

In this section we will apply our ideas to the bosonic string in $D$ dimensions. The Polyakov action is
\begin{equation}
\label{Poly} S_0=\frac{T}{2}\int d^2\sigma (\sqrt{-\gamma}\gamma^{ab}h_{ab}+b),
\end{equation}
where $\gamma_{ab}$ is the intrinsic metric 
\begin{equation}
h_{ab}=G_{\mu\nu}
\partial_a x^\mu
\partial_b x^\nu, 
\end{equation}
and the Kalb-Ramond NS-NS B field is defined through
\begin{equation}
b=\epsilon^{ab} B_{\mu\nu} 
\partial_a x^\mu 
\partial_b x^\nu.
\end{equation}
Here $G_{\mu\nu}$ is the space time metric. We define a first order action equivalent to the Polyakov action by
\begin{equation}
\label{Fo-string} S=\int d^2\sigma (\sqrt{-\gamma}\gamma^{ab}h_{ab}+b),
\end{equation}
where
\begin{equation}
h_{ab}=G_{\mu\nu}(-\frac{1}{2T}V^\mu_a V^\nu_b+V^\mu_a
\partial_b x^\nu),
\end{equation}
and
\begin{equation}
b=\epsilon^{ab} B_{\mu\nu}(-\frac{1}{2T} V_a^\mu V_b^\nu+  V_a^\mu\partial_b x^\nu),
\end{equation}
where $V_a^\mu$ are auxiliary fields. Notice that the intrinsic geometry is not changed in the first order action. This
imply that the symmetries of this first order action associated to the intrinsic geometry are the same as in the original
Polyakov action (\ref{Poly}). In particular
\begin{equation}
\delta \gamma_{ab}=\xi^c
\partial_c \gamma_{ab}+\gamma^{ac}
\partial_c\xi^b+\gamma^{bc}
\partial_c\xi^a,
\end{equation}
for diffeomorphisms in the world sheet and
\begin{equation}
\gamma_{ab}=\Omega \gamma_{ab},
\end{equation}
for the Weyl invariance. The space--time diffeomorphism invariance are easily extended to the first order action by
\begin{equation}
\delta x^\mu=\xi^c 
\partial_c x^\mu, \quad \delta V_a^\mu=
\partial_a(\xi^c V_c^\mu).
\end{equation}
To show the classical equivalence between the  action (\ref{Fo-string}) and  the Polyakov action (\ref{Poly}) lets
calculate the equation of motion associated with the auxiliary fields $V_a^\mu$
\begin{equation}
\frac{\delta L}{\delta V^\mu_a}=(\sqrt{-\gamma} \gamma^{ab} G_{\mu\nu}+\varepsilon^{ab}B_{\mu\nu})(\partial_b
x^\nu-\frac{1}{T} V^\nu_b)=0.
\end{equation}
From here we have
\begin{equation}
\label{V-def} V^\mu_a=T
\partial_a x^\mu.
\end{equation}
Using this equation to eliminate $V^\mu_a$ from  (\ref{Fo-string}) we obtain the original Polyakov action (\ref{Poly}) as desired.  

Now, to construct the dual theory we start from  the field redefinition \cite{duff}
\begin{equation}
\label{dual-1}
\varepsilon^{ab}
\partial_b X_\mu=(G_{\mu\nu}\sqrt{-\gamma}\gamma^{ab}+B_{\mu\nu}\varepsilon^{ab})\partial_b x^\nu.
\end{equation}
To extend this symmetry to the first order action we define
\begin{equation}
\label{dual-2} U^\mu_a=\frac{1}{T}V^\mu_a.
\end{equation}
We will call the transformation (\ref{dual-1}) and (\ref{dual-2}) a duality  transformation. The dual first order action
is then
\begin{equation}
S_D=T\int d^2\sigma (\sqrt{-\gamma}\gamma^{ab}h^D_{ab}+b^D),
\end{equation}
where
\begin{equation}
h^D_{ab}=\frac{1}{2}G_{\mu\nu}U^\mu_a U^\nu_b,
\end{equation}
and
\begin{equation}
b^D=\frac 12\epsilon^{ab} B_{\mu\nu} U_a^\mu U_b^\nu-U^\mu_a\varepsilon^{ab}\partial_b X_\mu .
\end{equation}
The equation of motion for the auxiliary field $U^\mu_a$ is
\begin{equation}
\frac{\delta L_D}{\delta U^\mu_a}=\sqrt{-\gamma}\gamma^{ab}U_b^\nu+\varepsilon^{ab}U_b^\nu B_{\mu\nu}-
\varepsilon^{ab}
\partial_bX_\mu,
\end{equation}
which imply
\begin{equation}
U^\mu_a=L^{\mu\nu}\frac{1}{\sqrt{-\gamma}}\gamma_{ab}\varepsilon^{bc}\partial_c X_\nu + M^{\mu\nu}\partial_aX_\nu.
\end{equation}
By the elimination of the auxiliary field $U^\mu$ from the dual first order action we obtain the dual Polyakov action given by
\begin{equation}
S_D=\int d^2\sigma (\sqrt{-\gamma}\gamma^{ab}h_{ab}+b),
\end{equation}
where
\begin{equation}\label{NM}
h_{ab}=L^{\mu\nu}
\partial_a X_\mu
\partial_b X_\nu,\quad 
b=\epsilon^{ab} M^{\mu\nu} 
\partial_a X_\mu 
\partial_b X_\nu,
\end{equation}
and 
$L=(G-BG^{-1}B)^{-1}$, $M=(B-gB^{-1}G)^{-1}$ are the space time metric and the Kalb-Ramond antisymmetric field
respectively. For future reference the inverse of the duality transformation is given by
\begin{equation}
\varepsilon^{ab}
\partial_b x^\mu=(L^{\mu\nu}\sqrt{-\gamma}\gamma^{ab}+M^{\mu\nu}\varepsilon^{ab})\partial_b X_\nu.
\end{equation}
With the duality transformation defined we proceed now to develop the Dirac formalism associated to the first order action
(\ref{Fo-string}).  As the intrinsic geometry of the string is not changed under the duality transformation
(\ref{dual-1},\ref{dual-2}) we can take the conformal gauge $\gamma_{ab}=\eta_{ab}$, without loss of generality. In this
gauge the constraints of the original theory can be written as
\begin{eqnarray}
\frac12 \delta^{ab}G_{\mu\nu}V_a^\mu V_b^\nu = 0,\\
 \frac12 \Omega^{ab}G_{\mu\nu}V_a^\mu V_b^\nu = 0,\\
 p^a_{\mu} = 0,\\
p_\mu+G_{\mu\nu}V_0^\nu-B_{\mu\nu}V_1^\nu = 0,\\
\partial_1 x^\mu-V_1^\mu =0,
\end{eqnarray}
where $p_\mu$ and $p^a_{\mu}$ are canonical variables associated with $x^\mu$ and $V_a^\mu$ respectively. The matrix
$\Omega^{ab}$ is the O(d,d,R) invariant
\begin{equation}
\Omega=
\begin{pmatrix} 0&1\cr 1&0
\end{pmatrix}.
\end{equation}
To split these constraints into first and second class constraints we need to redefine them. A consistent  redefinition is
\begin{eqnarray}
\nonumber {\cal H}= \frac12\left (\frac{1}{T}(p_\mu-B_{\mu\nu}\partial_\sigma x^\nu)^2+T (
\partial_\sigma x^\mu)^2\right)+\\
(G_{\mu\nu}V_0^\nu-B_{\mu\nu}V_1^\nu)
\partial_\sigma\chi^{2\mu} +(V_1^\mu)
\partial_\sigma\chi^1_\mu,\\
{\cal H}_1=p_\mu
\partial_\sigma x^\mu-(V_1^\mu)
\partial_\sigma\chi^2_\mu- (G_{\mu\nu}V_0^\nu-B_{\mu\nu}V_1^\nu)\partial_\sigma\chi^{1\mu},\\
\chi^1_\mu= p^0_\mu ,\\
\chi^2_\rho=B_{\mu\rho}G^{\mu\nu}p^0_\nu + p^1_\rho,\\
\chi^3_\mu=p_\mu+G_{\mu\nu}V_0^\nu-B_{\mu\nu}V_1^\nu ,\\
\chi_4^\mu=T
\partial_\sigma x^\mu-V_1^\mu ,
\end{eqnarray}
where ${\cal H}$ and ${\cal H}_1$ are first class and $\chi_i, i=1,2,3,4$ are second class. The nonzero Dirac brackets are 
\begin{eqnarray}
\{x^{\mu}(\sigma),V_{0}^{\nu}(\sigma')\}^{*}&=&
-G^{\mu\nu}\delta(\sigma-\sigma'),\\
\{V_{1}^{\mu}(\sigma),V_{0}^{\nu}(\sigma')\}^{*}&=&
-TG^{\mu\nu}\partial_{\sigma}\delta(\sigma-\sigma'),\\
\{p_{\mu}(\sigma),x^{\nu}(\sigma')\}^{*}&=&
-\delta^{\nu}_{\mu}\delta(\sigma-\sigma'),\\
\{p_{\mu}(\sigma),V_{0}^{\nu}(\sigma')\}^{*}&=&
TB_{\mu\lambda}G^{\lambda\nu}\partial_{\sigma}\delta(\sigma-\sigma'),\\
\{p_{\mu}(\sigma),V_{1}^{\nu}(\sigma')\}^{*}&=&
T\delta^{\nu}_{\mu}\partial_{\sigma}\delta(\sigma-\sigma'),
\end{eqnarray}
and the first class constraints algebra is 
$$
\{{\cal H}[\xi_1],{\cal H}[\xi_2]\}={\cal H}_1[\xi_1
\partial_\sigma\xi_2-\xi_2
\partial_\sigma\xi_1],$$
\begin{equation}\label{alg-cuerda}
\{{\cal H}[\xi_1],{\cal H}_1[\xi_2]\}={\cal H}[\xi_1
\partial_\sigma\xi_2-\xi_2
\partial_\sigma\xi_1],\end{equation}
$$
\{{\cal H}_1[\xi_1],{\cal H}_1[\xi_2]\}={\cal H}_1[\xi_1
\partial_\sigma\xi_2-\xi_2
\partial_\sigma\xi_1],$$
where ${\cal H}[\xi_i]$ denote the densitized constraint. This algebra is the standard algebra of the bosonic string theory \cite{bibi}.
If the metric $G_{\mu\nu}$ and the antisymmetric field $B_{\mu\nu}$ depend explicitly on the space-time coordinates
$X^\mu$ the complete algebra of constraints close in the same way as when the metric and the antisymmetric field are
independent of the space-time coordinates, up to second class constraints.

It is worth noticing that the correction terms added to the first class constraints ${\cal H}$ and ${\cal H}_1$ closes by
themselves as the first class algebra (\ref{alg-cuerda}). That means that they are a representation of the Virasoro algebra in twice the
number of space time dimensions. In addition they have the property of being linear in the momenta. 

To implement the duality transformation as a canonical transformation we extend, as in our previous example, the duality
transformation (\ref{dual-1}) to all phase space variables. The generator for this canonical transformation is
\begin{equation}\label{gen-1}
F_1(x,X)=T\int d\sigma x^\mu
\partial_\sigma X_\mu,
\end{equation}
while the rest of the variables transform as the identity map. This canonical transformation was constructed in \cite{p}
where the important observation that the transformation only works when the metric $G_{\mu\nu}$ and the antisymmetric
field $B_{\mu\nu}$ are independent of the space-time coordinates. If this is not the case the canonical transformation
generated by (\ref{gen-1}) will be nonlocal. Explicitly, the canonical transformation is
\begin{eqnarray}  \label{can-cuerda}
\nonumber p_\mu(\sigma)=\frac{\delta F_1(\sigma)}{\delta x^\mu(\sigma')}=T\partial_\sigma X_\mu(\sigma), &\quad&
P^\mu(\sigma)=-\frac{\delta F_1(\sigma)}{\delta X_\mu(\sigma')}=T
\partial_\sigma x^\mu(\sigma),\\
V_a^\mu=TU_a^\mu,&\quad& p^a_\mu=P^a_\mu,
\end{eqnarray}
where $X_\mu,  U_a^\mu, P^\mu, P^a_\mu $ are the new coordinates and its associated momenta. 

To construct the dual Dirac bracket the first step is to write down the dual constraint surface. The first class sector
remains essentially unchanged (up to an irrelevant factor of $T$ in the correction terms) and the dual second class
constraint surface is
\begin{eqnarray}
\Xi^1_\mu= P^0_\mu ,\\
\Xi^2_\rho=B_{\mu\rho}G^{\mu\nu}P^0_\nu + P^1_\rho,\\
\Xi^3_\mu=
\partial_\sigma X_\mu+G_{\mu\nu}U_0^\nu-B_{\mu\nu}U_1^\nu ,\\
\Xi_4^\mu=P^\mu-TU_1^\mu .
\end{eqnarray}
The nonzero Dual Dirac brackets are
\begin{eqnarray}
\{P^{\mu}(\sigma),U_{0}^{\nu}(\sigma')\}^{*}_D&=&
-G^{\mu\nu}\partial_\sigma\delta(\sigma-\sigma'),\\
\{U_{1}^{\mu}(\sigma),U_{0}^{\nu}(\sigma')\}^{*}_D&=&
-\frac1T G^{\mu\nu}\partial_{\sigma}\delta(\sigma-\sigma'),\\
\{X_{\mu}(\sigma),P^{\nu}(\sigma')\}^{*}_D&=&
\delta^{\nu}_{\mu}\partial_\sigma\delta(\sigma-\sigma'),\\
\{X_{\mu}(\sigma),U_{0}^{\nu}(\sigma')\}^{*}_D&=&
\frac1T B_{\mu\lambda}G^{\lambda\nu}\delta(\sigma-\sigma'),\\
\{X_{\mu}(\sigma),U_{1}^{\nu}(\sigma')\}^{*}_D&=&
\frac1T \delta^{\nu}_{\mu}\delta(\sigma-\sigma').
\end{eqnarray}

The implementation of the dual transformation by using the canonical transformation (\ref{can-cuerda}) is still non-local. Notice
that the canonical transformation does not depend explicitly in the space-time metric nor in the antisymmetric field
$B_{\mu\nu}$. Nevertheless, when we implement the reduction to the space defined by the variables $X_\mu, P^\nu$ the two
dual sigma models are related by a change in the space time metric and the antisymmetric field given by $L^{\mu\nu}$ and
$M^{\mu\nu}$ defined in (\ref{NM}).

\section{Conclusions}

Based on first order formalism we have found an implementation of S-Duality as a canonical transformation in a bigger phase space associated with the first order formulation. The effect of this symmetry transformation in the general analysis of the constrained dynamics and in particular in the second class sector revel that the Dirac bracket and the second class constraints change when we apply the symmetry map. The reduction of the action and the symmetry generator produces the well know results about canonical non local implementation of S-Duality where the first class sector was also solved. 

A different approach to S-duality as a symmetry of the action was constructed in \cite{dht, DT}. The price to pay for the symmetry is that the action does not have manifest Lorentz invariance. The analysis of linearized gravity \cite{HT, BCH, CMU} starting from a first order action can also be implemented using the ideas that we have worked here. It is also of interest to try to implement the S-duality as a symmetry of the action in theories that are not free \cite{BCHP, Julia}. In particular to analyze these works from the general perspective developed here.

 \section*{Acknowledgments}

This work was partially supported by Mexico's
National Council of Science and Technology (CONACyT), under grant
CONACyT-40745-F, and by DGAPA-UNAM, under grant IN104503-3.

\end{document}